\documentstyle[aas2pp4]{article}

\begin{document}
\title{EVN+MERLIN Observations of Radio-Intermediate Quasars: Evidence
for Boosted Radio-Weak Quasars}
\author{Heino Falcke}
\affil{Astronomy Department, University of Maryland, College Park,
MD 20742-2421 (hfalcke@astro.umd.edu)}
\author{Alok R. Patnaik (apatnaik@mpifr-bonn.mpg.de) \and William Sherwood (p166she@mpifr-bonn.mpg.de)}
\affil{Max-Planck-Institut f\"ur Radioastronomie, Auf dem H\"ugel 69,
D-53121 Bonn, Germany}
\vspace{0.7cm}

\centerline{\bf The Astrophysical Journal Letters, in press}

\begin{abstract}
We present VLBI (EVN+MERLIN) observations of a sample of three
low-redshift radio-intermediate PG quasars (RIQ) with flat and
variable radio spectrum (III~Zw~2, PG 1309+355, PG 2209+184).  Their
radio-to-optical flux ratio ($R$) is slightly lower than the average
$R$ for steep-spectrum quasars, but their radio spectral properties are
those of core-dominated quasars. It was proposed previously that these
sources might be relativistically boosted jets in radio-weak quasars.
Our VLBI observations now indeed confirm the presence of a high
brightness temperature core in all three of these objects --- two of
them have lower limits on $T_{\rm B}$ well in excess of $10^{10}$
Kelvin. Moreover, we find no ``missing-flux'' which means that
basically all the flux of these quasars is concentrated in the compact
radio core. As the total radio flux is already at the low end for
radio-loud quasars, we can place a strong limit on the presence of any
extended emission.  This limit is consistent with the extended
emission in radio-weak quasars, but excludes that the flat-spectrum
RIQ reside in typical radio-loud quasars. The observations therefore
strongly support the idea that relativistic jets are present in
radio-weak quasars and hence that radio-loud and radio-weak quasars
have very similar central engines.
\end{abstract}


\section{Introduction}
The radio properties of quasars with otherwise very similar optical
properties can be markedly different. As Strittmatter et al.~(1980)
and Kellerman et al.~(1989) showed, there is a dichotomy between
radio-loud and radio-weak quasars which cannot be explained by a
single orientation-based scheme.  Radio-loudness is ususally defined
by the $R$-parameter --- the radio-to-optical flux ratio,
e.g.~Kellermann et al.~(1989) suggested that $R\sim10$ separates
radio-loud from radio-weak quasars. Investigations of the radio
morphology of PG quasars (Miller, Rawlings, \& Saunders 1993 and
Kellermann et al.~1994) show that there are, in fact, three main types
of radio structures associated with quasars: flat-spectrum
core-dominated sources, steep-spectrum lobe-dominated sources and
those with weak diffuse emission. In the optically selected PG quasar
sample (see Schmidt \& Green 1983), all of the steep-spectrum
radio-loud quasars have a core-lobe structure, which, in morphology
and flux, is similar to those of FR\,II radio galaxies, while
radio-weak quasars have only weak and diffuse extended emisson. For
the radio-loud quasars this is expected within the basic unified
scheme (e.g.~Barthel 1989) where it is generally assumed that radio
galaxies are just quasars seen edge on such that the optical nucleus
is hidden from our line of sight. Moreover, for the compact
core-dominated sources Orr and Browne (1982) already suggested that
they might simply be the face-on counterparts to radio-loud quasars
where relativistic boosting in a radio jet enhances the core flux, and
swamps the steep-spectrum and largely orientation-independent
lobe-emission. One of the critical tests for the unification of
core-dominated and lobe-dominated quasars was the investigation of
their extended emisson. And indeed Browne \& Perley (1986) and Murphy,
Browne, \& Perley (1993) found that the extended emission of
core-dominated quasars and BL Lacs is compatible with those of radio
galaxies.

Thus, it seems as if the radio-loudness in quasars -- be they core or
lobe dominated -- can be attributed to the existence of a powerful,
relativistic jet, while radio-weak quasars seemingly have a different
engine. However, recently Falcke, Sherwood, \& Patnaik (1996,
hereinafter FSP96) investigated the radio spectral indices and average
$R$-parameters of PG quasars and found that the fraction of compact
flat-spectrum sources ($\sim40\%$) in the PG sample with $R>10$ is far
larger than expected in the unified scheme for radio-loud
quasars. Moreover, the median $R$-parameter of flat-spectrum
radio-loud quasars was {\em lower} than those of steep-spectrum
radio-loud quasars.  As the extended steep-spectrum lobes should
radiate largely isotropically, it appears impossible that the parent
population of all radio-loud flat-spectrum sources are steep-spectrum
radio-loud quasars. Instead FSP96 suggested that there is a population
of flat-spectrum radio-intermediate quasars (RIQ, see also Miller,
Rawlings, \& Saunders 1993, and Falcke, Malkan, Biermann 1995) which
could be relativistically boosted {\em radio-weak} quasars. FSP96
showed that moderate Lorentz factors of 2-4 in radio-weak quasars are
sufficient to boost the radio fluxes at 5 GHz of a substantial number
of radio-weak quasars into the regime $R>10$ usually occupied by
(un-boosted) steep-spectrum radio-loud quasars.  Wilson \& Willis
(1980) had already suggested earlier that some core-dominated Seyferts could
have relativistically boosted jet. The question whether there is
relativistic boosting in radio-weak quasars can therefore be quite
important for the proper classification of quasars and for the
explanation of the radio-loud/radio-quiet dichotomy.

A crucial test for the ``boosted radio-weak jet'' hypothesis are
high-resolution VLBI observations of the RIQ. If they are indeed
boosted relativistic jets, one expects high brightness temperatures
and possibly core-jet structures, and if they are to be preferentially
oriented radio-weak quasars they should -- unlike radio-loud quasars
-- be naked cores on higher resolution images, i.e.~having only very
weak extended emission. On the other hand, if the RIQ fail this test
and are not related to radio-weak quasars, one would have a strong
argument that radio-weak quasars are {\em not} relativistic in their
cores at all, because in the PG quasar sample are no sources left, besides
the flat-spectrum RIQ, that are unaccounted for
and could be the boosted counterparts to radio-weak quasars.

The criteria to select flat-spectrum RIQ are that their spectral index
$\alpha$ should be larger than $-0.5$ ($S_\nu\propto\nu^\alpha$), and
their $R$-parameter should be larger than 10 (the usual threshold for
a radio-loud quasar) and smaller than 250 (the median $R$-parameter
for steep-spectrum quasars). To reduce the effects of variability it
is helpful to use a time averaged radio-flux for each source as given
for example in FSP96.

In this Letter we now report VLBI (EVN+MERLIN) observations of the
three low-redshift ($z<0.2$) flat-spectrum RIQ in the PG quasar sample
that satisfy the above criteria. Those sources are: PG 0007+106
(III~Zw~2), PG 1309+355, and PG 2209+184.

\section{Observations and Data Reduction}

Simultaneous observations were made with 6 telescopes of the
European-VLBI-Network (EVN) and MERLIN on 1995 May 26-27 at 5 GHz
using a bandwidth of 28 MHz (Mk3 VLBI recording in mode B). Each
source was observed for about 4 hours. 3C286 was used as the primary
flux density calibrator for MERLIN observations and the EVN data were
calibrated using system temperature measurements. Both of the datasets
were analyzed separately using the NRAO AIPS software package and the
Caltech DIFMAP package.

\section{Results}
In Fig.~1 we show the maps produced from our data.  All maps are
restored with a circular gaussian of 5 mas FWHM and natural
weighting. The size of the maps are 512 $\times$ 512 pixels with a
pixel size of 1.5 mas $\times$ 1.5 mas. The flux densities are
expected to have an accuracy of 5 to 10\% for III Zw 2 and PG
2209+184.  The noise levels are 0.3 mJy/beam and 0.6 mJy/beam for III
Zw 2 and PG 2209+184 respectively.  If we compare our MERLIN fluxes
--- which are within the errors similar to the VLBI fluxes --- with
the VLA observations by Kellermann et al.~(1989) we can confirm the
strong variability in all sources --- while the variability in III Zw
2 is well established (e.g.~Ter\"asranta et al.~1992), PG 2209+184 has
now a flux 2 times higher, and PG 1309+355 has a flux three times
lower than in the earlier VLA measurement.

None of our sources was resolved and we did not find any significant
extended structure in the EVN or the MERLIN data. The sources are also
unresolved in the VLA A\&D array maps by Kellermann et
al.~(1994). In order to appear unresolved in our maps the sources have
to be even smaller than our nominal beam size of 5 mas. The visibility
function for the longest baseline and the shortest baseline for all
sources are indentical, which translates into an upper limit of
$\sim$1 mas for the source sizes. At least for the two brighter
sources, a somewhat less conservative estimate for our high
Signal-to-Noise data could actually be a factor 2 lower, i.e. 0.5
mas. Using the formula

\begin{equation}
T_{\rm B}=1.8\cdot10^9\,{\rm K}\; {S_{\nu}\over{\rm mJy}}
\left({\nu\over{\rm GHz}}{{\rm FWHM}\over{\rm mas}}\right)^{-2}
\end{equation}
we then calculated a lower limit for the brightness temperatures of
these sources, which for III Zw 2 and PG 2209+184 are in excess of
$10^{10}$ Kelvin and may almost approach $10^{11}$ Kelvin if the less
conservative size limit is used.

Moreover, by comparing the total flux density to that obtained in the
EVN and MERLIN maps, we found no evidence for any missing flux at a
5\% level in the two brighter sources and at the 30\% level in PG
1309+355. This gives us a strong limit for the flux from any extended
radio components, e.g.~as expected in radio-galaxies. If we divide
this upper limit to the extended flux by the optical flux at 4400\AA{}
as given in Kellermann et al.~(1989), we obtain an upper limit to the
$R$-parameter for the extended flux only ($R_{\rm ext}$) --- this
value should be much less subject to orientation effects. The limits
for $R_{\rm ext}$ obtained this way\footnote{For III Zw 2 Unger et
al.~(1987), detected a week and steep second component with the VLA (8
mJy at 1490 MHz, i.e. $\sim3.2$ mJy at 5 GHz for
$\alpha={-.75}$). Likewise, Kellermann et al.~(1994) claim a $\sim$ 2
mJy secondary component for PG 2209. If these week components contain
all the extended flux, one would obtain an even lower value of $R_{\rm
ext}\sim2$ for III Zw 2 and PG 2209.} for our sample of RIQ are listed
in Tab.~1 and are all smaller than $R_{\rm ext}$=10. We note that for
the steep-spectrum radio-loud quasars always $R_{\rm ext}\simeq R$ and
for radio-weak quasars with detected extended emission $R_{\rm ext}$
also is only slightly less than $R$ (see Kellermann et al.~1989). In
Fig.~2 we show the distribution of $R$ (reproduced from FSP96) and
$R_{\rm ext}$ for the PG quasars, where we have left out all
flat-spectrum sources except for the three RIQ investigated here, to
highlight how the RIQ's classifications change if one considers the
extended rather than the total emisson. While in total flux the three
flat-spectrum RIQ are in the radio-loud distribution, the upper limits
for their extended emssion push them already down to the tip of the
radio-weak and below the radio-loud distribution.

\begin{deluxetable}{lrrrrrrrrr}
\small
\tablecaption{Properties of low-redshift flat-spectrum RIQ}
\tablehead{
\colhead{Name}&
\colhead{$z$}&
\colhead{$\left<\alpha_{\rm 5\,GHz}\right>$}&
\colhead{$\left<S_{\rm 5\,GHz}\right>$}&
\colhead{$\left<R\right>$}&
\colhead{$S_\nu({\rm core})$}&
\colhead{FWHM}&
\colhead{$S_\nu({\rm ext})$}&
\colhead{$R_{\rm ext}$}&
\colhead{$T_{\rm B}$}\\
\colhead{}&
\colhead{}&
\colhead{}&
\colhead{[mJy]}&
\colhead{}&
\colhead{[mJy]}&
\colhead{[mas]}&
\colhead{[mJy]}&
\colhead{}&
\colhead{[K]}
}
\tablecolumns{10}
\startdata  
PG 0007$+$106&0.089&   0.66&158&258&$234\pm12$&$<$1&$<12$&$<7  $&$>1.7\cdot10^{10}$\nl
PG 1309$+$355&0.184&$-$0.02& 34& 11&$ 15\pm\hphantom{1}5$&$<$1&$<5$&$<1.7$&$>1.1\cdot10^{9\hphantom{0}}$\nl
PG 2209$+$184&0.070&   0.32& 92&188&$255\pm13$&$<$1&$<13$&$<6  $&$>1.8\cdot10^{10}$\nl
\tablecomments{The columns are 
(1) -- source name, 
(2) -- redshift,
(3) -- time averaged spectral index,
(4) -- time averaged total radio flux at 5 GHz,
(5) -- time averaged radio-to-optical flux ratio (all three from FSP96),
(6) -- VLBI core flux,
(7) -- upper limit to FWHM of source size in order to appear
unresolved in our map,
(8) -- missing VLBI flux,
(9) -- upper limit for radio-to-optical flux ratio of extended flux,
(10) -- lower limit for core brightness temperature
}
\enddata
\end{deluxetable}

\section{Summary and Discussion}
The hypothesis that the flat-spectrum RIQ in the low-redshift PG
quasar sample are actually boosted radio-weak quasars was initially
based only on their peculiar position in an radio vs.~optical diagram
and their flat radio spectrum (Falcke, Malkan, \& Biermann 1995, see
also Miller et al.~1993). The notion was then further strengthened by
the finding of more flat-spectrum RIQ in the whole PG quasar sample
(FSP96), which in number and $R$ distribution were then consistent
with the boosting hypothesis. Our VLBI observations of the three
low-redshift RIQ in the PG sample now confirm two of the three
predictions for these quasars in more detail: a high $T_{\rm B}$ and a
low extended flux. The lower limits for brightness temperatures of
$10^{10}$ Kelvin for the two brighter quasars are already large enough
to exclude any reasonable thermal models (e.g. the starburst model),
especially in conjunction with their strong variability. If we take
the less conservative estimates for the source sizes, the limits on
the brightness temperatures for III~Zw~2 and PG 2209+184 already
approach the theoretical limit of $\sim10^{11}$ Kelvin derived from
equipartition arguments (e.g.~Falcke \& Biermann 1995, Eq.~58) which
would imply relativistic boosting. For III Zw 2 Ter\"asranta \&
Valtaoja (1994) even estimate a $T_{\rm B}$ around $10^{12}$ K from
their variability data. Such high values for $T_{\rm B}$ are typical
for compact radio cores in radio galaxies and radio-loud quasars
associated with powerful radio jets.  However, the upper limit on the
extended flux of the flat-spectrum RIQ we have obtained reduces the
already low $R$-parameter to values at $R_{\rm ext}<10$ --- values,
which are typical for radio-weak quasars. This basically excludes that
these compact, variable, high-brightness cores reside in the usual
type of radio-loud quasar with bright extended emission from the
large-scale lobes, otherwise we would have expected a missing flux of
at least 30-50\%, based on the typical lobe fluxes of the
steep-spectrum radio-loud quasars --- this was neither seen in our EVN
+ MERLIN maps nor in the earlier VLA A\&D array maps. The third
prediction of the ``boosted radio-weak jet'' hypothesis, namely
core-jet structure and superluminal motion could not be tested in our
experiment because of our relatively low resolution. For this we have
to await future VLBA or even global VLBI experiments. Nevertheless,
our observations show that the RIQ are indeed a homogenous class of
sources with unique properties and they have brought strong, direct
evidence for the presence of relativistic boosting in radio-weak
quasars.

This finding may be quite significant for our interpretation of
radio-weak quasars and especially the radio-loud/radio-weak dichotomy.
It means that radio-weak and radio-loud quasars have central engines
that are in many respects very similar: not only are the optical
properties almost undistinguishable but also do {\em both} types of
quasars produce relativistic jets in their nuclei. Unfortunately, this
makes the reason for the radio dichtomy of quasars an even deeper
mystery.

\acknowledgements This research was supported in part by NASA under
grants NAGW-3268 and NAG8-1027.  MERLIN is a U.K. national facility
operated by the University of Manchester on behalf of PPARC. We thank
the staff of the participating EVN observatories and of the MPIfR
correlator for their assistance. We have benefited from many
interesting disucssions with A. Wilson and P. Biermann on this and
related topics.

\clearpage

\onecolumn
\clearpage
\begin{figure*}
\epsscale{0.32}
\centerline{\plotone{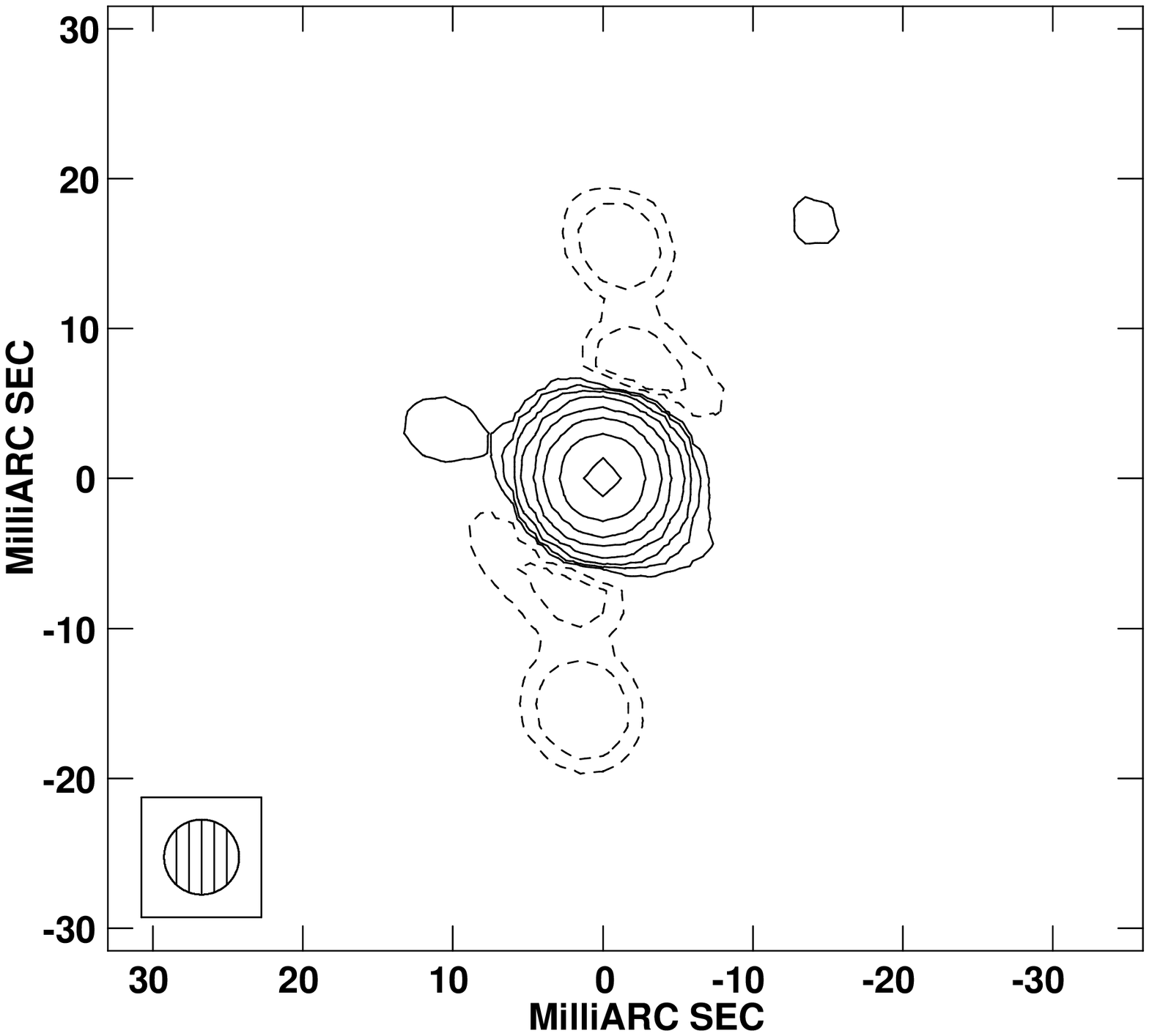}\plotone{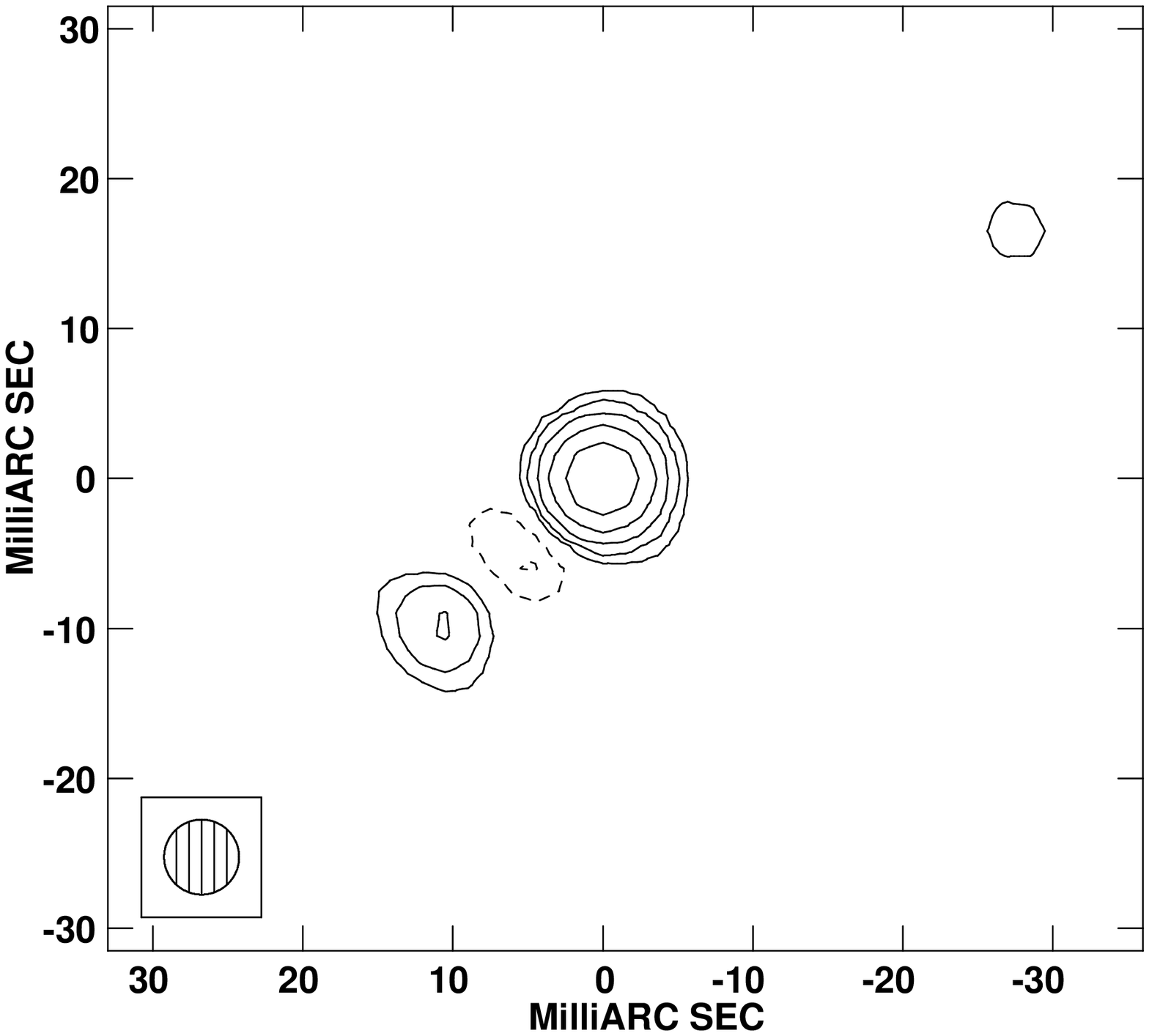}\plotone{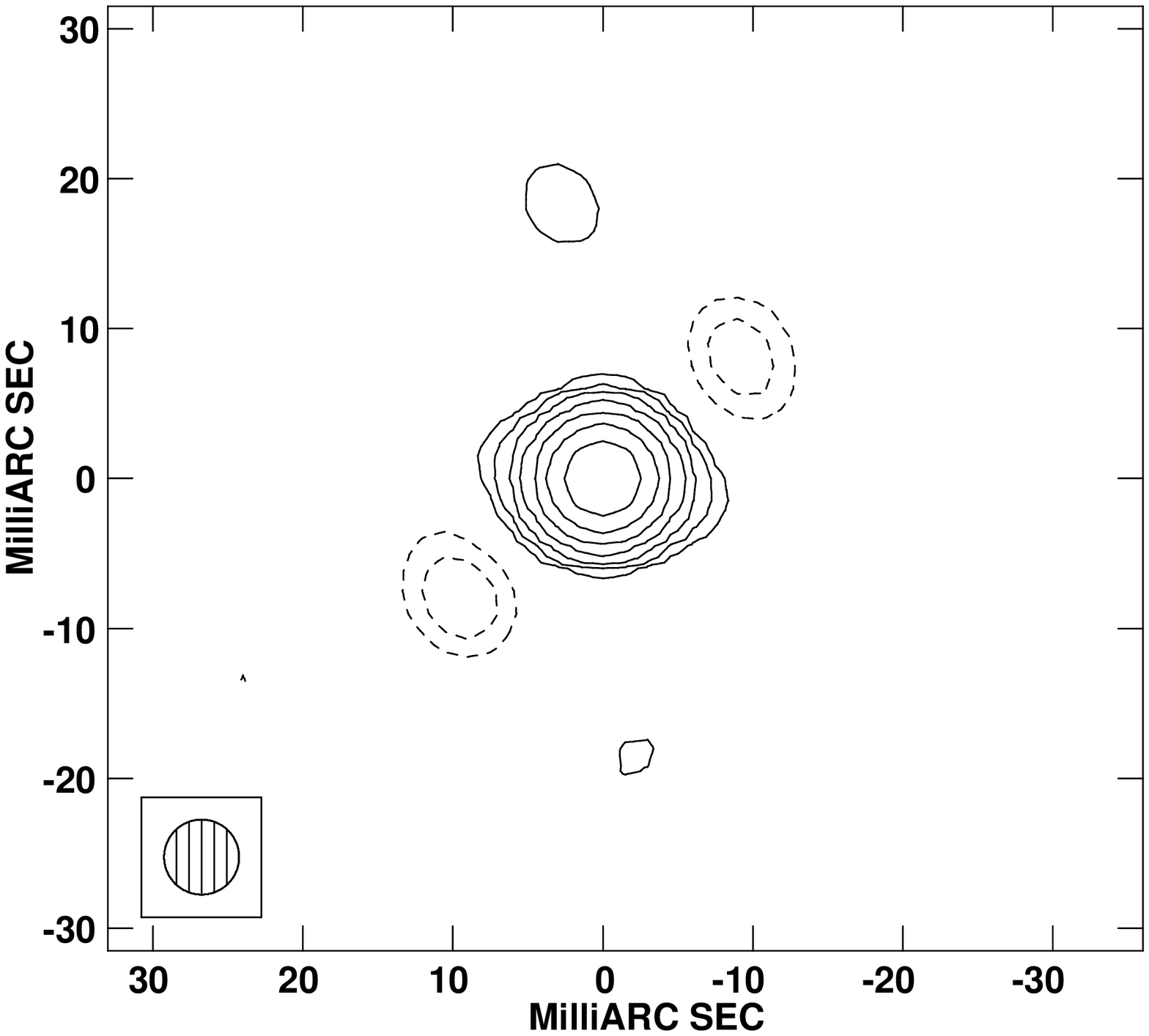}}
\caption[]{VLBI maps at 5 GHz of the three low-redshift RIQ in the
PG quasar sample. From left to right: III~Zw~2, PG~1309+355, and
PG~2209+184. The maps were restored with a beam size of 5 mas and the
noise is less than 0.6 mJy per beam. All sources are unresolved. The
secondary component in PG 1309+355 is too weak to be considered
significant. Contour levels are
(-2,-1,1,2,4,8,16,32,64,128,256)$\times1.5$ mJy, and peak fluxes are
given in Tab. 1.}
\end{figure*}
\clearpage
\begin{figure*}
\epsscale{0.6}
\centerline{\plotone{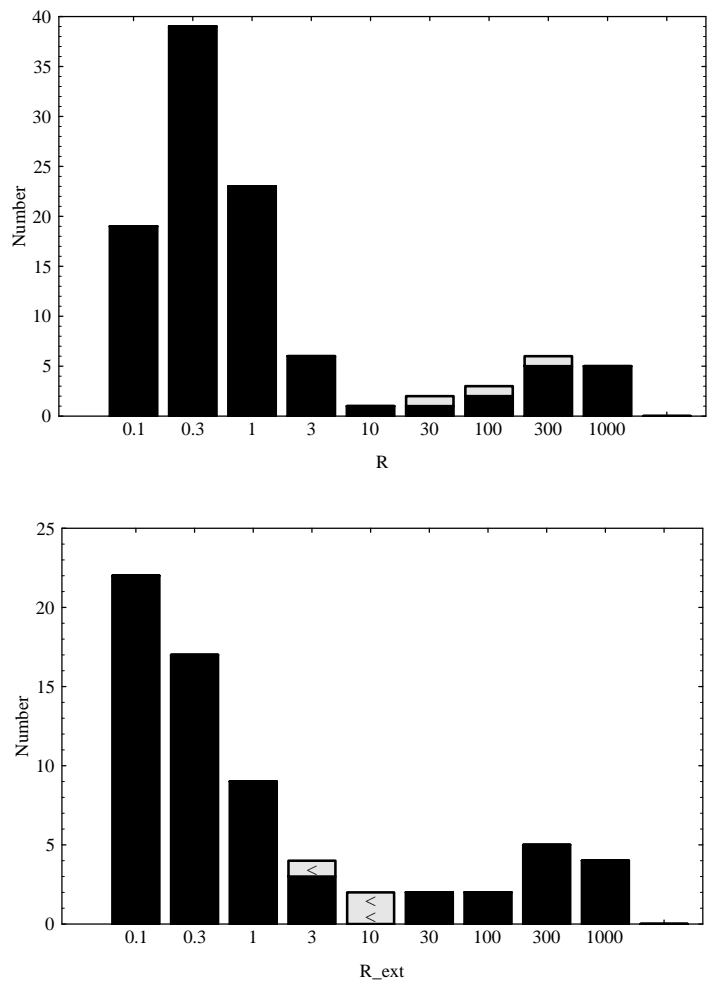}}
\caption[]{Top: The distribution of the (total)radio-to-optical
flux ratio for ``steep-spectrum'' PG quasars (i.e.~all sources except
those with known flat spectrum --- the spectral information is only
complete down to $R\sim1$) is shown as filled bars (from FSP96). The
position of the three flat-spectrum RIQ investigated in this paper are
shown as shaded bars.\hfill\break 
Bottom: Distribution of the extended-radio-to-optical flux ratio
$R_{\rm ext}$ for all ``steep-spectrum'' PG quasars with extended
emission (black bars) and the three flat-spectrum RIQ (shaded bars),
where we have used the upper-limits derived from this paper. While in
total flux the flat-spectrum RIQ appear to be part of the radio-loud
distribution, their low limits on the extended flux indicate that they
might rather be part of the radio-weak distribution.}
\end{figure*}

\end{document}